\documentclass[%
 reprint,
preprintnumbers,
 amsmath,amssymb,
 aps,
nofootinbib
]{revtex4-1}

\usepackage[normalem]{ulem} 
\usepackage{graphicx}  
\usepackage{color}
\usepackage{dcolumn}   
\usepackage{bm}        
\usepackage{hyperref}  

\graphicspath{
{./}
  {./media/}
}

\def\ka{\kappa}
\def\la{\lambda}

\def\si{\sigma}

\newcommand{\ben}{\begin{equation}}
\newcommand{\een}{\end{equation}}
\newcommand{\bea}{\begin{eqnarray}}
\newcommand{\eea}{\end{eqnarray}}
\newcommand{\ba}{\begin{array}}
\newcommand{\ea}{\end{array}}
\newcommand{\bal}{\begin{align}}
\newcommand{\eal}{\end{align}}
\newcommand{\bit}{\begin{itemize}}
\newcommand{\eit}{\end{itemize}}

\newcommand{\dBV}{d_\text{BV}}

\newcommand{\vrmss}{\bar{v}_\text{s}}
\newcommand{\vrmsm}{\bar{v}_\text{m}}

\newcommand{\vrel}{\bar v_\text{rel}}
\newcommand{\vperp}{\bar v_{\parallel}}
\newcommand{\xis}{\xi_\text{s}}
\newcommand{\xim}{\xi_\text{m}}

\newcommand{\mMon}{M_\text{m}}

\newcommand{\tauCG}{\tau_\text{cg}}

\begin{document}

\preprint{HIP-2018-11/TH}

\title{Scaling in necklaces of monopoles and semipoles}

\newcommand{\Sussex}{\affiliation{
Department of Physics and Astronomy,
University of Sussex, Falmer, Brighton BN1 9QH,
U.K.}}

\newcommand{\HIPetc}{\affiliation{
Department of Physics and Helsinki Institute of Physics,
PL 64, 
FI-00014 University of Helsinki,
Finland
}}

\author{Mark Hindmarsh}
\email{m.b.hindmarsh@sussex.ac.uk}
\Sussex
\HIPetc

\author{Anna Kormu}
\email{anna.kormu@helsinki.fi}
\HIPetc

\author{Asier Lopez-Eiguren}
\email{asier.lopezeiguren@helsinki.fi}
\HIPetc

\author{David J. Weir}
\email{david.weir@helsinki.fi}
\HIPetc

\date{\today}

\begin{abstract}
  Models of symmetry breaking in the early universe can produce
  networks of cosmic strings threading 't Hooft-Polyakov
  monopoles. In certain cases there is a larger global symmetry group
  and the monopoles split into so-called semipoles. These networks are
  all known as cosmic necklaces. We carry out large-scale field theory
  simulations of the simplest model containing these objects,
  confirming that the energy density of networks of cosmic necklaces
  approaches scaling, i.e. that it remains a constant fraction of the
  background energy density. The number of monopoles per unit comoving
  string length is constant, meaning that the density fraction of
  monopoles decreases with time. Where the necklaces carry semipoles
  rather than monopoles, we perform the first simulations large enough
  to demonstrate that they also maintain a constant number per unit
  comoving string length.  We also compare our results to a number of
  analytical models of cosmic necklaces, finding that none explains
  our results.  We put forward evidence that annihilation of poles on
  the strings is controlled by a diffusive process, a possibility not
  considered before. The observational constraints derived in our
  previous work for necklaces with monopoles can now be safely applied
  to those with semipoles as well.
\end{abstract}

\maketitle

\section{Introduction}

Symmetry-breaking phase transitions in the early universe are a
natural consequence of attempts to explain physics beyond the Standard
Model, for example by incorporating the elements of the Standard Model
in a Grand Unified Theory (GUT). Depending on the nature of the
symmetry that is broken during such a phase transition, it is possible
for topological defects to have formed.  Defects are solitonic
solutions of the field equations carrying conserved topological
charge; however, the word is used more loosely to mean any extended
classical structures in the field, including long-wavelength Goldstone
modes.

In cosmology the most interesting defects are cosmic strings
\cite{Kibble:1976sj} (see
Refs.~\cite{Hindmarsh:1994re,Vilenkin:2000jqa,Copeland:2011dx,Hindmarsh:2011qj}
for reviews).  They appear even in the simplest case of the Abelian
Higgs model, forming when the $\mathrm{U}(1)$ gauge symmetry
breaks. The cosmic strings arising from this symmetry breaking are
Nielsen-Olesen vortex lines~\cite{Nielsen:1973cs}. Similar objects can
also arise as fundamental objects in an underlying string
theory. These objects, termed F- and D-strings, are also known as
cosmic
superstrings~\cite{Witten:1985fp,Sarangi:2002yt,Copeland:2003bj,Urrestilla:2007yw,Lizarraga:2016hpd}.

More complex patterns of symmetry-breaking or models with extra
dimensions can produce structures which are combinations of different
kinds of defect.  Models of this type that have attracted attention in
recent years include semilocal
strings~\cite{Vachaspati:1991dz,Achucarro:1999it,Achucarro:2005tu,Achucarro:2013mga,Lopez-Eiguren:2017ucu},
which are a combination of Goldstone modes and cosmic strings;
necklaces~\cite{Hindmarsh:1985xc,Aryal:1987sn,Berezinsky:1997td,Tong:2003pz,Ng:2008mp,Kibble:2015twa,Hindmarsh:2016lhy},
which are a combination of strings and monopoles; and related models
where the monopoles form string
junctions~\cite{Vachaspati:1986cc,Hindmarsh:2006qn}.  The first direct
numerical simulations of necklace networks were performed by some of
the authors of this article in Ref.~\cite{Hindmarsh:2016dha}.

In this paper we continue our investigation of the non-Abelian strings
started in Ref.~\cite{Hindmarsh:2016dha}. The particular theory with
which we work models, in its most basic form, a two-stage GUT
symmetry-breaking scenario where first an $\mathrm{SU}(2)$ symmetry
breaks to a $\mathrm{U}(1)$, forming 't Hooft-Polyakov
monopoles. Later, at a lower symmetry-breaking scale, this
$\mathrm{U}(1)$ itself breaks in a manner analogous to that in the
simpler Abelian Higgs model. The result is that the magnetic flux of
the 't Hooft-Polyakov monopoles is then carried by two cosmic string
segments linking the monopoles together. This, then, spontaneously
breaks a $Z_2$ symmetry that relates the magnetic charge of the
monopoles and the orientation of the strings.

A monopole that is attached to two cosmic strings in this way is
termed a `bead' and a system of many such beads on a loop of string
forms a `necklace'. As shown in Ref.~\cite{Hindmarsh:2016lhy} the
`beads' can be seen as the kinks form when $Z_2 \times Z_2$ symmetry
is spontaneously broken to $Z_ 2$ by the string solutions. An exact
solution is known in a model with ${\cal N} = 2$ supersymmetry
\cite{Tong:2003pz}.

When the symmetry-breaking scales are degenerate, the global symmetry
$Z_2 \times Z_2$ is enlarged to $D_4$, the square symmetry group.
When $D_4$ breaks down to $Z_2$, the kinks that are formed are labeled
by a $Z_4$ topological charge and they can be seen as split
beads. That is, each bead separates into two `semipoles'.  Semipoles
can annihilate only with the corresponding anti-semipole. Unlike
monopoles, two adjacent semipoles need not have total charge zero, and
they can repel each other \cite{Hindmarsh:2016lhy}. We will refer to
monopoles and semipoles collectively as `poles'.  Semipoles come in
two different types, depending on a ratio of dimensionless couplings
$\la$ and $\ka$ in the scalar potential \cite{Hindmarsh:2016lhy}. In
Ref.~\cite{Hindmarsh:2016dha} we simulated only with $\ka/2\la \ge 1$;
here, we perform the first simulations for $\ka/2\la < 1$.  We do not
revisit the special case $\kappa/2\lambda = 1$, where the symmetry
group on the string is enhanced to $\mathrm{O}(2)$ and semipoles do
not exist.

In order to characterise the gross features of a network of cosmic
necklaces we can use two length scales: the average comoving pole
separation, $\xim$, and the average comoving string separation,
$\xis$. These quantities are of great interest for the analysis of the
network evolution because they show if the system has reached
scaling. Scaling is a important property for the reliable study of
defect networks, because it tells us how to extrapolate network
observables to large cosmic times.  Scaling, in its simplest form
  as applied to cosmic string networks, means that all quantities
with dimensions of length grow in proportion to the horizon distance,
$\xi \propto \tau$.  In a scaling string network, the fraction of the
energy density coming from defects remains constant.  However,
necklaces have an important dynamical length scale
\cite{Berezinsky:1997td}
\begin{equation}
\label{e:dBVdef}
\dBV = \frac{\mMon}{\mu},
\end{equation}
where $\mMon$ is the monopole mass and $\mu$ the string mass per unit
length.\footnote{The mass of a monopole or semipole on a string is
  generally less than that of a free pole, but still the same order of
  magnitude.}  The inverse $1/\dBV$ sets the scale for the
acceleration of a monopole attached to a bent string.  For strings
alone, the local acceleration is equal to the curvature, so there is
no fixed scale in the dynamics. This is the underlying reason for why
strings approach scaling. One cannot apply the same argument to
necklaces, and their scaling is more difficult to understand.

It turns out to be informative to study the linear comoving monopole density
\begin{equation}
n = \frac{\xis^2}{\xim^3},
\label{eq:n}
\end{equation}
or equivalently the linear physical monopole density in units of
$\dBV$ \cite{Berezinsky:1997td},
\begin{equation}
\label{e:rDef}
r = \dBV n/a,
\end{equation}
where $a$ is the cosmological scale factor.

The mean comoving energy density of the network is 
\begin{equation}
\rho_n \simeq \frac{\mu}{\xis^2}(1+r),
\label{eq:xin}
\end{equation}
from which one can see that $r$ is the string-to-monopole mean energy
density ratio.  Therefore, if the strings scale ($\xis \propto \tau$)
and $r$ is a constant, the network will maintain a constant density
fraction.

One would expect that when $r\ll 1$ the string evolves essentially
without regard to the poles.  On the other hand, when $r$ is
significant the evolution of the network should change in some way.

Firmly establishing the behaviour of $r$, or equivalently $n$, is
important for predictions of observable signals from necklaces,
including the production of high energy cosmic rays, cosmic microwave
background fluctuations, and gravitational waves.

In Ref.~\cite{Berezinsky:1997td}, it was suggested that the density of
monopoles on strings would grow to be so large as to dominate the
dynamics. This would slow the string network down, leading to large
numbers of monopole-antimonopole annihilation events and a copious
source of ultra-high energy cosmic rays.

On the other hand, Ref.~\cite{BlancoPillado:2007zr} argued that
monopoles acquire substantial velocities along the string, similar in
magnitude to the transverse velocities of the strings themselves,
leading to frequent monopole interaction events on the string, and
efficient monopole annihilation.  The number of monopoles per unit
length should therefore decrease towards the minimum allowed by
causality $1/t$, and the strings should end up behaving like an
ordinary cosmic string network, with RMS velocity a significant
fraction of the speed of light.

In Ref.~\cite{Martins:2010ma} the velocity-dependent one-scale model
was adapted to necklace models, with the principal conclusion being
that both $\xis$ and $\xim$ should be expected to scale in most
circumstances, and that the monopole velocities are driven towards
unity, with continuously increasing Lorentz factors.

With contradictory results from analytical studies, direct numerical
simulations are required. 
In Ref.~\cite{Hindmarsh:2016dha}, we carried
out the first field-theory simulations of the system, but with
restricted dynamic range the conclusions we could draw were rather
limited.
Evidence was presented that the
monopole-necklace system evolves towards a state with a linear
increase in the comoving string separation $\xis$ with conformal time
and $r$ tending to zero in such a way that $n$ remained approximately
constant.  In the semipole case, $n$ appeared to increase towards the
end of the simulations.  The behaviour was not definitively
established as the simulations were not large enough. In all cases,
the energy density of the necklaces was transferred efficiently to
propagating modes of the gauge and scalar fields, much as for Abelian
Higgs cosmic strings~\cite{Hindmarsh:2017qff}, implying that necklaces
are not an important source of gravitational waves.

In the present paper we go beyond these earlier simulations, and
establish firmly the scaling properties of the network.  We are able
to analyse larger mass ratios than before. We also explore the effect
of different defect separations in the initial conditions.

We are able to reject important hypotheses made in the previous
model-building attempts outlined above, in particular: the
monopole-to-string density ratio $r$ never increases, in contradiction
with the Berezinky-Vilenkin model~\cite{Berezinsky:1997td}; the
decrease is slower than $r \propto t^{-1}$, in
contradiction with the Blanco-Pillado and Olum
model~\cite{BlancoPillado:2007zr}; and the monopole velocities
asymptote to a constant value, in contradiction with the Martins
model~\cite{Martins:2010ma}.

We confirm that the monopoles pick up a substantial component of
velocity along the string \cite{BlancoPillado:2007zr}.  We also
confirm the findings of Ref.~\cite{Hindmarsh:2016dha} that the scaling
state for the string-monopole system has a linear increase in the
comoving string separation $\xis$ with conformal time, and constant
comoving linear monopole density $n$.  

For necklaces with semipoles we find similar behaviour, independent of
the parameter ratio $\ka/2\la$ which controls their type: like
necklaces with monopoles, both the RMS velocity and the comoving
linear density $n$ tend to a constant. 

We have been unable to produce a satisfactory model that explains the
observed monopole and semipole densities. The fact that the monopole
density decreases more slowly than envisaged in the model of
Blanco-Pillado and Olum means that monopole annihilation is not as
efficient as proposed, but we have not been able to establish why. We
put forward a proposal based on pole diffusion in the Discussion.

The paper is organised as follows: In Sections \ref{sec:model} and
\ref{sec:num} we describe the model and the numerical
simulations. Then in Section~\ref{sec:res} we show the results
obtained and in Section~\ref{sec:ana} we compare them to necklace
evolution models. Finally, in Section~\ref{sec:dis} we discuss the
results obtained.

\section{Model}
\label{sec:model}

The model that we study is the $\mathrm{SU}(2)$ Georgi-Glasgow model
with two Higgs fields in a spatially flat Robertson-Walker metric. In
this section we will introduce the model and summarise its most
important aspects. A more detailed description of the model can be
found in Refs. \cite{Hindmarsh:1985xc,Hindmarsh:2016lhy}.

In comoving coordinates $x^i$, conformal time $\tau = x^0$, and with
scale factor $a$, the action is
\begin{eqnarray}
\mathcal{S}=\int d^4x \Bigg( -\frac{1}{4}F^a_{\mu\nu}F^{\mu\nu a}+a^ 2 \sum_n \mathrm{Tr} [D_{\mu},\Phi_n] [D^{\mu},\Phi_n] \nonumber \\ 
-a^ 4 V(\Phi_1,\Phi_2) \Bigg), \nonumber \\
\end{eqnarray}
where $D_{\mu}=\partial_{\mu}+igA_{\mu}$ is the covariant derivative,
$A_{\mu}=A_{\mu}^a\si^a/2$, and $\sigma^a$ are Pauli matrices.  The
Higgs fields $\Phi_n$, $n=1,2$, are in the adjoint representation,
$\Phi_n=\phi^a_n\si^a/2$.  Spacetime indices have been raised with the
Minkowski metric with mostly negative signature.

The potential can be written in the following way:
\begin{eqnarray}
V(\Phi_1,\Phi_2) &=& - m_1^2\mathrm{Tr}\Phi_1^2 - m_2^2\mathrm{Tr}\Phi_2^2  \nonumber\\
&+&\lambda (\mathrm{Tr}\Phi_1^2)^2 + \lambda (\mathrm{Tr}\Phi_2^2)^2 +\kappa(\mathrm{Tr}\Phi_1\Phi_2)^2,
\label{eq:potential}
\end{eqnarray}
where $\lambda$ and $\kappa$ are positive and $m_{1,2}$ are real. 

The system undergoes two symmetry-breaking phase transitions,
$\mathrm{SU}(2)\rightarrow \mathrm{U}(1) \rightarrow Z_2$. After the first
symmetry-breaking the theory has 't Hooft-Polyakov monopole solutions
and after the second one the theory has string solutions. The vacuum
expectation values of the two adjoint scalar fields are given by
$\mathrm{Tr}\Phi^2_{1,2} = m^2_{1,2}/2\lambda$, where the scalar
masses are then $\sqrt{2}m_{1,2}$. Without loss of generality we will
take that $\Phi_1$ has the larger vacuum expectation value, that is,
it is the responsible field for the first symmetry-breaking.

Depending on the value of the parameters of the potential, $m_1$,
$m_2$, $\lambda$ and $\kappa$, the model can accommodate three
different kinds of solutions, see Ref.~\cite{Hindmarsh:2016lhy}:

\begin{itemize}
\item When $m_1^2>m_2^2$ the system has a discrete global $Z_2 \times
  Z_2$ symmetry under which $\Phi_1\rightarrow \pm \Phi_1$ and
  $\Phi_2\rightarrow \pm \Phi_2$. The string solutions break $Z_2
  \times Z_2$ to $Z_2$ and the resulting kinks are the beads that
  interpolate between two string solutions. This solutions can be
  interpreted as 't Hooft-Polyakov monopoles with their flux confined
  to two tubes.

\item When $m_1^2=m_2^2$ the system has a square symmetry $D_4$ which
  is broken to $Z_2$ by strings. The resulting kinks can be seen as
  beads that are split into two. Each one of these kinks are known as
  semipoles. Semipoles can only be annihilated with the corresponding
  anti-semipole. Two classes of solutions exist according to whether
  $\ka/2\la <1$ or $\ka/2\la > 1$.

\item When $m_1^2=m_2^2$ and $\kappa/2 \lambda=1 $ there is a global
  $\mathrm{O}(2)$ symmetry. This symmetry is spontaneously broken by
  the string solution but not the vacuum. In this case there are no
  semipoles and the strings carry persistent global currents. We do
  not investigate this case here.
\end{itemize}

\section{Simulation Details}
\label{sec:num}

\subsection{Numerical Setup}

We discretise the system on a comoving 3D spatial lattice with lattice
spacing of $dx=1$ and time-step of $d\tau=0.1$. Then, the lattice
equations of motion are evolved using the standard leapfrog method. We
perform $1920^3$ simulations in the radiation dominated era, for which
$a \propto \tau^\nu$ with $\nu =1$.  More information about the
discretisation and simulation details can be found in
Ref.~\cite{Hindmarsh:2016dha}.

Analysis of observables should be done once the system has reached
scaling, so that extrapolation to cosmologically relevant times is
possible. To reduce uncertainties, we want scaling to be reached over
as large a time interval as possible, and this can be achieved by
choosing a `good' set of initial conditions. The aim is to generate a
random distribution of well-separated defects with otherwise minimal
field excitations, consistent with the field configuration expected at
a large time after the phase transition. The details of the phase
transition itself are not important for the late-time field
configuration.

In our case we choose $\Phi_{1,2}$ to have uniformly distributed
random values in the range $[-0.5, 0.5]$ for each component
$\phi^a_{1,2}$, which we then normalise to the vev of the field in
question. The $\mathrm{SU}(2)$ gauge field is set up by generating a
random $\mathrm{SU}(2)$ matrix from four Gaussian random numbers
${u^0,u^a}$ which are normalised to obtain a unitary matrix of
determinant 1.

Once the initial field configuration is set we smooth the
configuration of the Higgs fields, that is, in each lattice point we
substitute the field value by a weighted average of the field values
at the actual lattice point and at the six nearest neighbours:
\begin{equation}
\Phi_n(\mathbf{x}) \to \frac{1}{12} \sum_i \left[
  \Phi_n(\mathbf{x}-\hat{\imath}) + 2\Phi_n(\mathbf{x}) +
  \Phi_n(\mathbf{x} + \hat{\imath}) \right].
\end{equation}
We apply this smoothing $N_s$ times to the initial configuration, a
number which is in general different for the two fields. The aim of
this differential smoothing is to explore networks with different
initial densities of monopoles and strings, allowing us to vary $n$.

After smoothing the initial configuration we run with relatively
strong damping period for a time $\delta \tau_d$. The damping term is
handled using the Crank-Nicolson method~\cite{Crank1996}, but is
rather stronger than adopted in Ref.~\cite{Hindmarsh:2016dha}; we take
$\sigma=4$ in the notation of that paper.

The heavy damping phase ends at $\tau=120$, after which we run the
simulation with the standard Hubble damping for one light-crossing
time of the box, at which point the conformal time is $\tau_\text{end}
= 2040$. 

As with all simulations in fixed comoving volume in an expanding
background, physical widths such as the size of the defects shrink,
which presents a two-fold problem: to make sure they are
well-separated in the beginning and well-resolved at the end.  A
common approach in field theory simulations of this type is to scale
the couplings and mass parameters with factors $a^{1-s}$, where $a$ is
the cosmological scale factor and $0 \leq s \leq 1$. This procedure
keeps the scalar expectation values fixed and the string tension
constant but the comoving width of the string core grows for $s<1$. In
our simulations we use $s=1$, but we run with $s=-1$ from the end of
the damping period until time $\tau_{cg}$. This means that the
comoving width of the string can be made small while they are formed.
It also accelerates the production of the network, because the
conformal time taken by the fields to settle to their vacua is of
the order the comoving defect width. The scale factor is normalised to
$a(\tau_\text{end}) = 1$, so that the defects remain resolved
throughout the simulation.

In principle, correlations can start to be established after half a
light-crossing time. However, the only massless excitations are waves
on the string, and the strings are much longer than the box size even
at the end of the simulations.  We therefore do not expect finite-size
effects, although we check for small deviations from scaling towards
the end of the simulations.

\subsection{Measurements}
\label{sec:measure}

During the simulation we measure the number of poles $N$ and the
string length $L$. In order to obtain the monopole number we compute
the magnetic charge in each lattice site. The string length is
computed by counting the plaquettes pierced by strings, that is,
counting the plaquettes with a gauge-invariant `winding' in the
$\mathrm{U}(1)$ subgroups formed by projection with the scalar field
$\Phi_1$, the heavier one in the non-degenerate case.

In the case of monopoles, this measurement process gives the magnetic
field $\mathbf{B}^{(1)}$ and, by calculating the divergence, the exact
number of monopoles. For semipoles with $\kappa/2\lambda < 1$, this
yields approximately half the semipoles; the rest are sources or sinks of a magnetic
field $\mathbf{B}^{(2)}$ obtained by projecting out the
$\mathrm{U}(1)$ gauge field associated with
$\Phi_2$~\cite{Hindmarsh:2016lhy}. Finally, when $\kappa/2\lambda >
1$, the relevant magnetic fields are $\mathbf{B}^{(\pm)} =
(\mathbf{B}^{(1)} \pm \mathbf{B}^{(2)})/\sqrt{2}$, and our measurement
of $\mathbf{B}^{(1)}$ sources and sinks picks out features in the
field configuration of a string rather than the semipoles
themselves. We call these midpoints `pseudopoles'.

On the other hand, the measurements of the winding number -- and hence
the string length and velocity -- do not depend on the particular
choice of projecting scalar field. See the Appendix of
Ref.~\cite{Hindmarsh:2016dha} for details of the projectors used.

\begin{figure} 
  \includegraphics[width=3.5in]{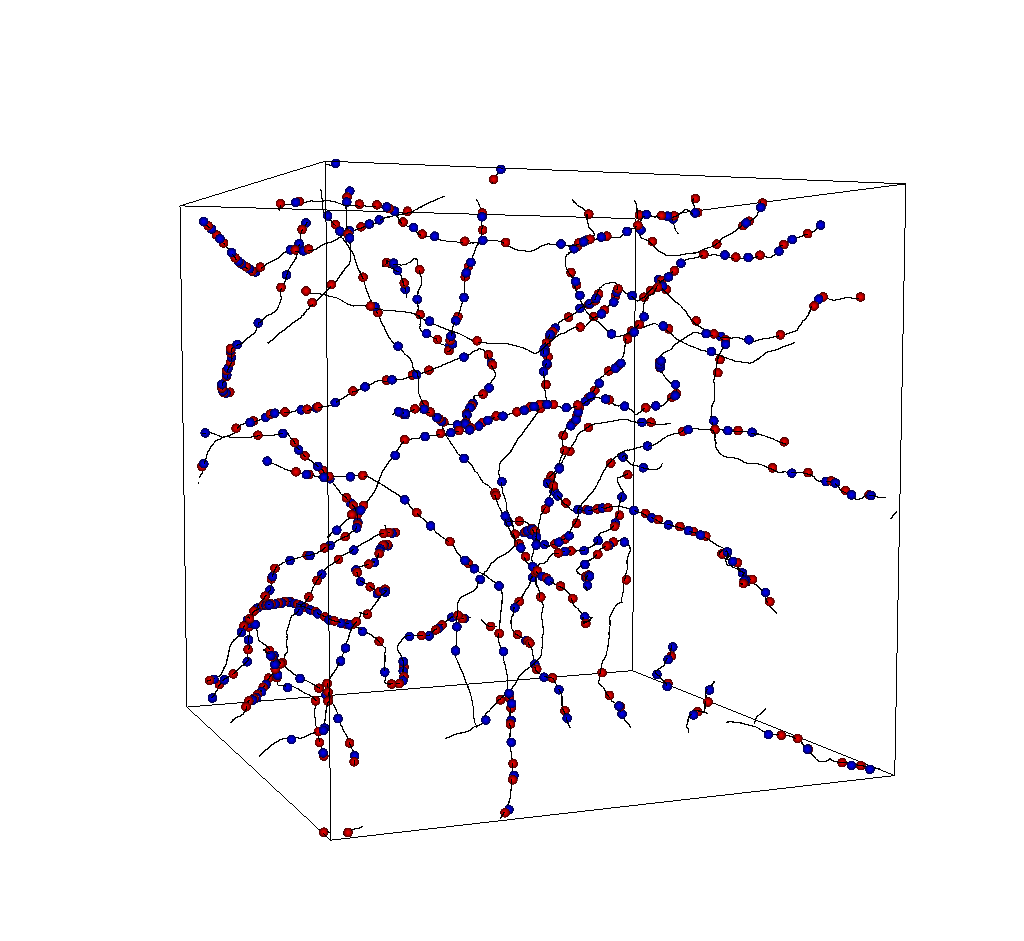}
   \caption{Snapshot of a semipole necklace at the end of the
     simulation. The black lines represent the strings, the red
     circles the poles picked out by $\mathbf{B}^{(+)}$ and the blue circles the poles picked out by $\mathbf{B}^{(-)}$. The run parameters are given in the last entry
     of Table \ref{tab:param}.}
   \label{fig:snap}
\end{figure}

Fig.~\ref{fig:snap} shows a snapshot of the end of one of the semipole
simulations (the next-to-last in Table \ref{tab:param}), with the strings in
black and the semipoles represented by red and blue circles.

Using the pole number $N$, and the string length $L$ we can derive the
average comoving defect separations as
\begin{equation}
\xim=(V/N)^{1/3}, \quad \xis=(V/L)^{1/2},  
\label{separations}
\end{equation}
from which we calculate the linear comoving pole density [Eq.~(\ref{eq:n})]
and $r$ [Eq.~(\ref{e:rDef})]. As explained above, the quantity $N$ for
semipoles and pseudopoles counts only half the total number, but we
retain the definition as it is more directly comparable with the
number of monopoles.

We use the positions of the strings and poles to compute the string
root-mean-square (RMS) velocity $\bar{v}_s$, and the monopole RMS
velocity $\bar{v}_m$. Once we have the positions of poles and strings
at each time step we can follow their trajectories during the
simulation. Computing the trajectory at every time step $d\tau$
is computationally very expensive, and it can also induce some
noise due to lattice discretisation ambiguities.  Therefore, we
perform the computations to obtain the trajectories every time
interval $\delta \tau_v = 20 \; d\tau$.

We also record global quantities such as the total energy and
pressure, from which energy conservation can be checked.  In all runs
global covariant energy conservation is maintained to 1\% or better.
Detailed information about the measurements can be found in
Ref.~\cite{Hindmarsh:2016dha}.

\subsection{Parameter choices}

We analyse the cases with degenerate $(m_1^2=m_2^2)$ and
non-degenerate mass parameters, which allows us to study both
monopoles (discrete global $Z_2 \times Z_2$ symmetry) and semipoles
(square $D_4$ symmetry).  In the semipole case we analyse two
different parameter relations, $\kappa/2 \lambda > 1$ and (for the
first time) $\kappa/2 \lambda < 1$.

For monopoles, we explore various ratios of $m_1^2$ to $m_2^2$ and
various initial configurations, that is, different values for the
smoothing iterations, $N_s$ and different damping periods $\delta
\tau_d$. More precisely, all the runs are carried out with
$m_1^2=0.25$ in the radiation-dominated era ($\nu=1$) and the scale
factor is normalised so that $a=1$ at the end of the simulations.

The values of the rest of the parameters can be seen in
Table~\ref{tab:param}. We perform one realisation for each set of
parameter choices.

\begin{table}
\centering
\scalebox{0.95}{
\begin{tabular}{ccccccccccc}
$m_1^2$ & $m_2^2$ & $\lambda$ & $\kappa$ & $M_m$ & $\mu$ & $d_{\mathrm{BV}}$ & $N_s$ & $\delta \tau_d$  & $\tau_{cg}$ & $t_{cg}$ \\ \hline
0.25 & 0.1 & 0.5 & 1 & 11 & 0.63 & 17.5 & 10000/10000 & 350 & 520  & 66\\
0.25 & 0.1 & 0.5 & 1 & 11 & 0.63 & 17.5 & 4000/10000 & 350 & 520 & 66\\
0.25 & 0.1 & 0.5 & 1 & 11 & 0.63 & 17.5 & 1000/10000 & 350 & 520 & 66\\
0.25 & 0.1 & 0.5 & 1 & 11 & 0.63 & 17.5 & 4000/10000 & 87.5  & 520 & 66\\ \hline
0.25 & 0.025 & 0.5 & 1 & 11 & 0.16 & 70 & 4000/4000 & 350 & 520 & 66 \\
0.25 & 0.0125 & 0.5 & 1 & 11 & 0.08 & 140 & 4000/4000 & 350   & 520 & 66\\ \hline
0.25 & 0.25 & 0.5 & 0.25 & 11 & 1.6 & 7 & 4000/4000 & 350 & 520 & 66\\
0.25 & 0.25 & 0.5 & 0.5 & 11 & 1.6 & 7 & 4000/4000 & 350  & 520 & 66\\
0.25 & 0.25 & 0.5 & 2 & 11 & 1.6 & 7 & 4000/4000 & 350 & 520 & 66\\
0.25 & 0.25 & 0.5 & 4 & 11 & 1.6 & 7 & 4000/4000 & 350 & 520 & 66\\
\hline
\end{tabular}
}
\caption{List of simulation parameters for the runs we performed. The
  dimensionful parameters are given in units of the lattice spacing
  $dx$. Potential parameters (\ref{eq:potential}) are shown
  along with the isolated monopole mass $\mMon$ and the isolated
  string tension $\mu$. The length scale $\dBV=\mMon/\mu$ is also
  shown as well as the smoothing iterations $N_s(\Phi_1)/N_s(\Phi_2$),
  damping time $\delta \tau_d$ and end of the core growth period in
  conformal time, $\tau_{cg}$, and physical time,
  $t_{cg}$. Simulations were run until conformal time $\tau = 2040$.}
\label{tab:param}
\end{table}

\section{Results \label{sec:res}}

\subsection{Length Scales \label{sec:lensc}}

\begin{figure} 
   \centering \includegraphics[width=3.5in]{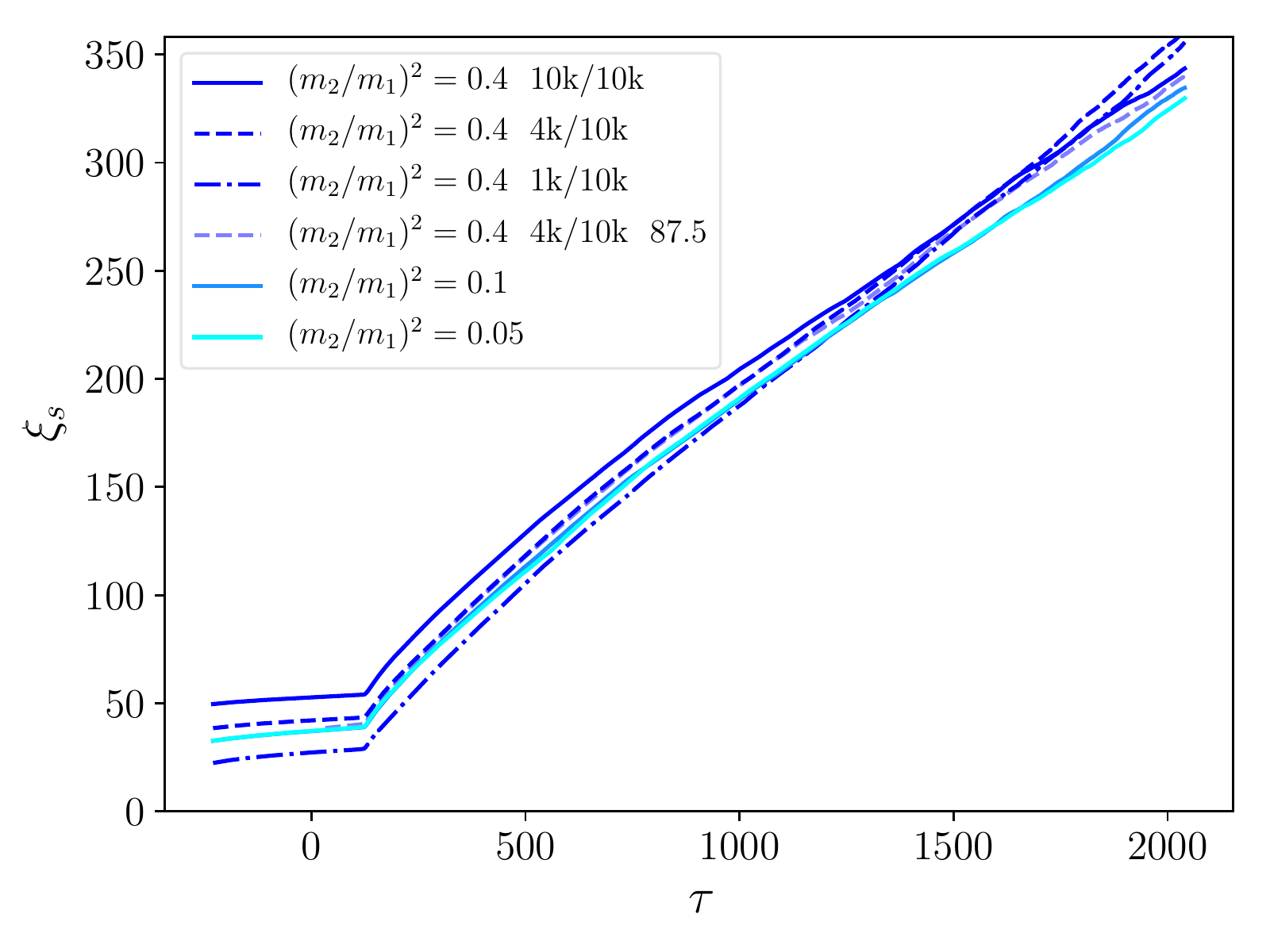}
   \includegraphics[width=3.5in]{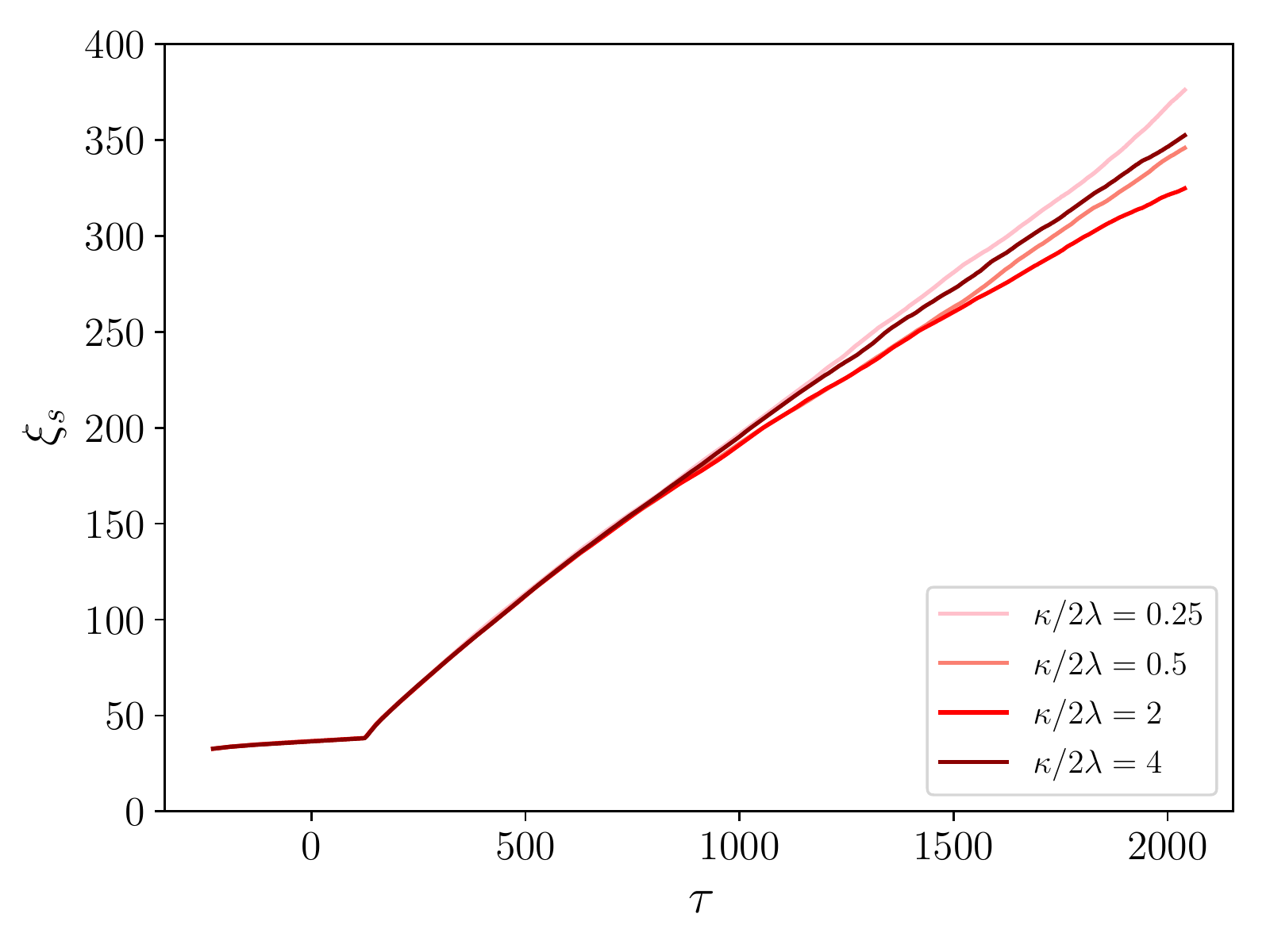}
   \caption{Mean string separation $\xis$, defined in
     Eq.~(\ref{separations}), for necklaces with monopoles (top) and
     semipoles (bottom), against conformal time $\tau$.  The legend
     gives the ratio mass-squared values of the fields $(m_2/m_1)^2$
     for the necklaces with monopoles and the ratio of scalar couplings $\kappa/2\la$ for
     the necklaces with semipoles. In the case where the mass-squared
     ratio is $0.4$ the legend also shows the number of smoothing
     steps performed in each field as $N_s(\Phi_1)/N_s(\Phi_2)$.  We
     distinguish the case with the equal amount of smoothing showing
     the damping time $\delta \tau_d$ where it is the shortest.  A
     full list of simulation parameters is given in Table
     \ref{tab:param}.  }
   \label{fig:xis}
\end{figure}

\begin{figure} 
   \centering
   \includegraphics[width=3.5in]{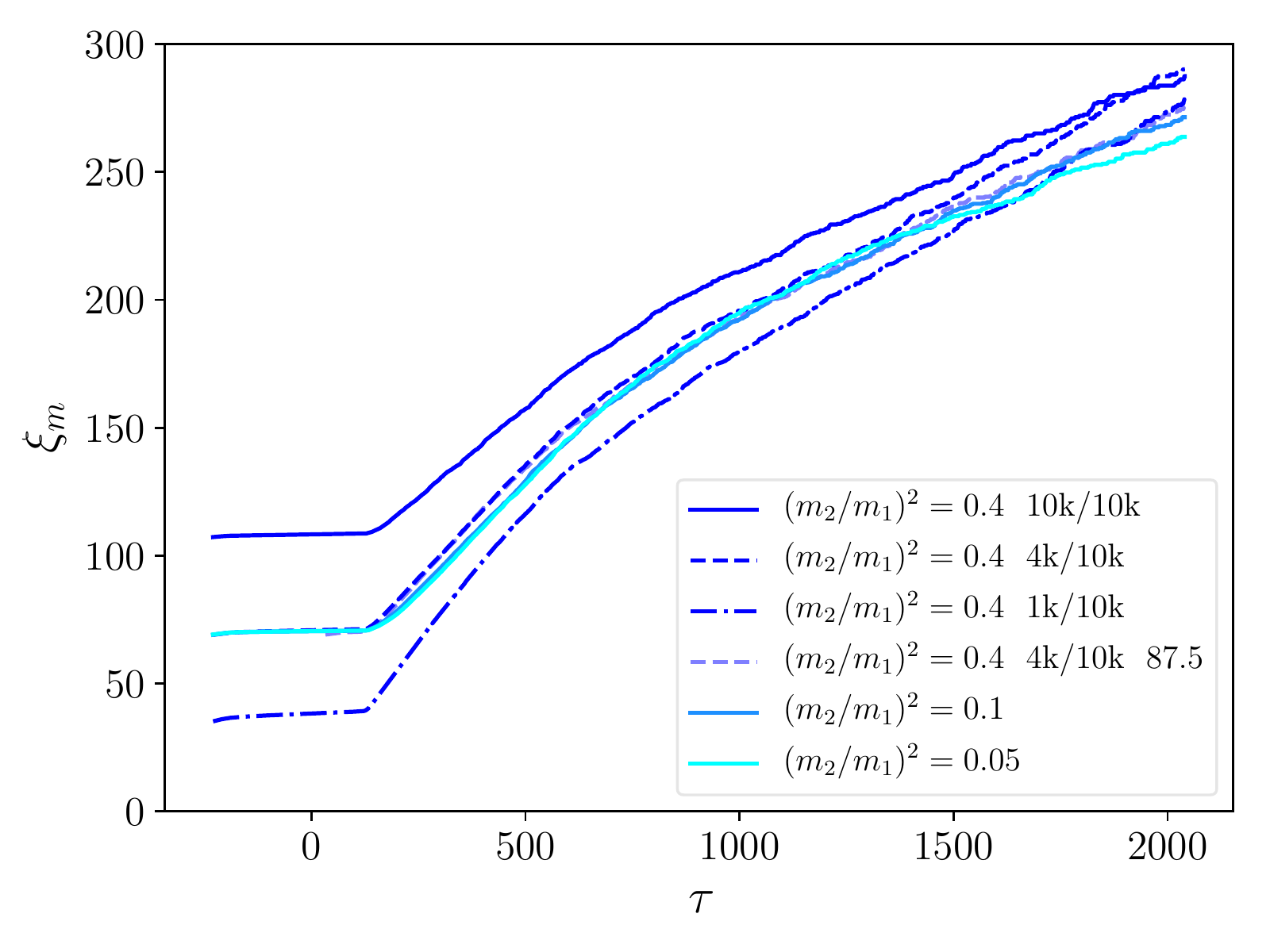} 
   \includegraphics[width=3.5in]{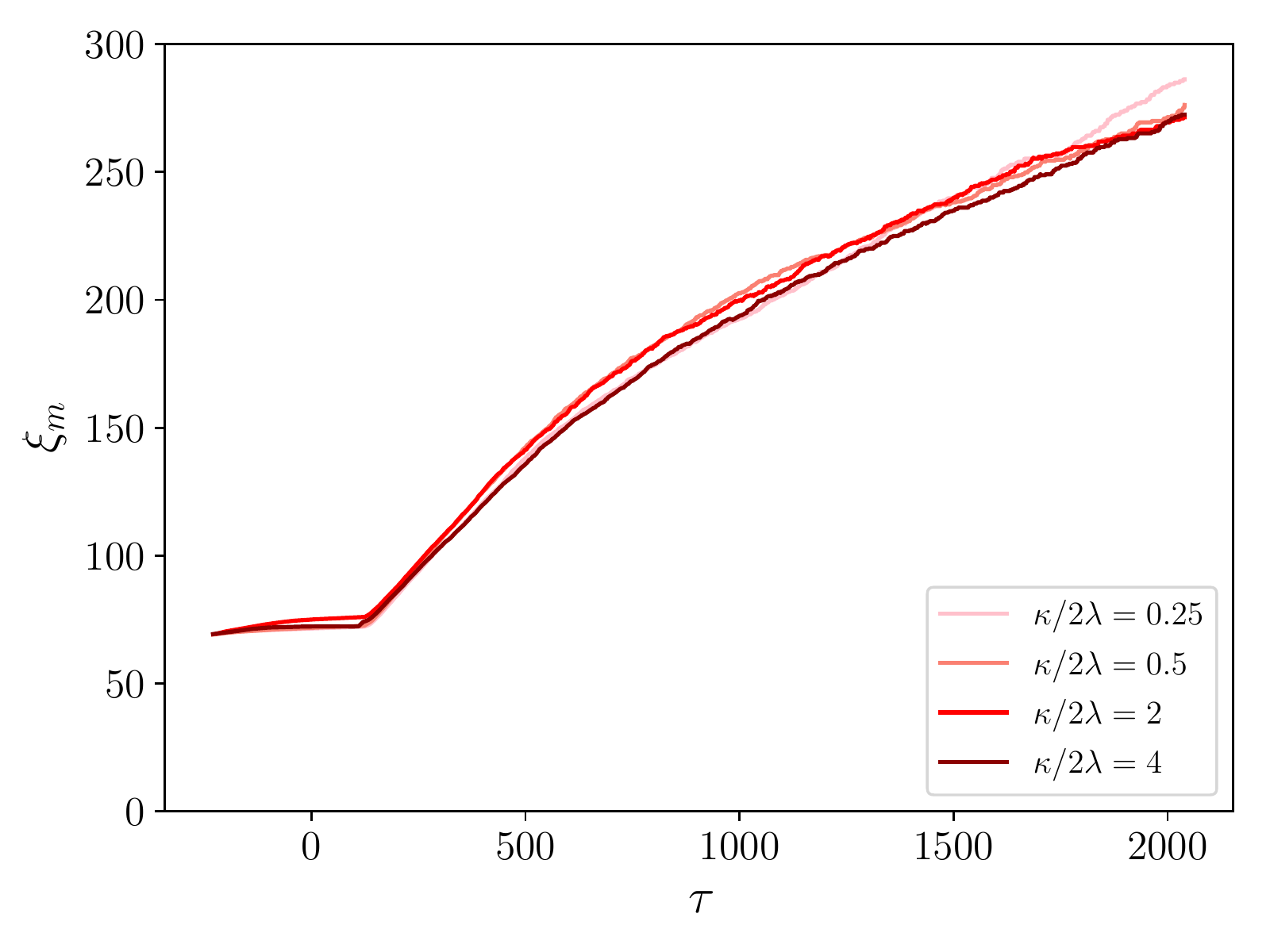} 
   \caption{Mean monopole separation $\xim$, defined in
     Eq.~(\ref{eq:xin}), for necklaces with monopoles (top) and
     semipoles (bottom), against conformal time $\tau$. See the
     caption to Fig.~\ref{fig:xis} for an explanation of the legend.}
   \label{fig:xim}
\end{figure}

The comoving necklace network length scales $\xis$ and $\xim$, which
are defined in Eq.~(\ref{separations}) are plotted in
Figs.~\ref{fig:xis} and~\ref{fig:xim}. In these plots we show all the
cases for which we have carried out simulations.

The effect of the different amounts of smoothing in the initial
conditions can bee seen in the initial defect separations: the more
smoothing, the further apart the defects. The amount of damping makes
little difference to the initial defect separation, but does reduce
the oscillations in the RMS deviation of the field from its vacuum
value, $\delta\Phi_{1,2} = |\mathrm{Tr}\Phi^2_{1,2} -
v_{1,2}^2|^{1/2}$. The subsequent evolution depends little on the
initial conditions: the system evolves towards a scaling regime
characterised by $\xis \propto \tau$.

In order to analyse the scaling regime we have computed the gradients
for the comoving string separation $\xis$, in three different time
regimes. These time ranges, which are $\tau \in [1000, 1250]$, $\tau
\in [1250, 1500]$ and $\tau \in [1500, 1750]$, are chosen to cover the
biggest part of the dynamical range taking into account that the
system needs some time to reach scaling after the core growth
period. The values of the gradients can be seen in
Table~\ref{tab:xi}. The gradients confirm that the strings are indeed
scaling. In addition, we can conclude that the finite-size effects are
negligible, because the values of the gradients at the final time
range are compatible with the values at the other two time ranges.

\begin{table}
  \centering
  \begin{tabular}{ccccccc}
$m_2^2/m_1^2$ &$\kappa/2\lambda$ & $(\frac{d\xis}{d\tau})_1$ & $(\frac{d \xis}{d\tau})_2$ &$(\frac{d\xis}{d\tau})_3$ & Mean & Std   \\[2pt] \hline
0.4 & 1 & 0.131 & 0.140 &  0.138   & 0.136  & 0.005  \\
0.4 & 1 & 0.147 & 0.153 &  0.152   &  0.151 & 0.003   \\
0.4 & 1 & 0.154 & 0.162 &   0.151   & 0.156 &  0.006   \\
0.4 & 1 & 0.134 & 0.155  &  0.135   & 0.141 & 0.012  \\ \hline
0.1 & 1 & 0.139 & 0.132 & 0.134  & 0.135 & 0.004   \\
0.05 & 1 & 0.140 & 0.134 & 0.125 & 0.133 & 0.008    \\ \hline
1 & 0.25 & 0.167 & 0.166 & 0.156 & 0.163  & 0.006   \\
1 & 0.5 & 0.136 & 0.148 & 0.162 & 0.149 & 0.013  \\
1 & 2 & 0.139  & 0.140 & 0.127  &  0.135 &  0.007   \\
1 & 4 & 0.157 & 0.154 &  0.151 & 0.154 & 0.003  \\
\hline
\end{tabular}
  \caption{Results of the $\xi_s$ gradients computed in three
    different ranges. Numerical annotations refer to the range in
    which the gradient is computed: 1 has $\tau \in [1000, 1250]$, 2
    has $\tau \in [1250, 1500]$ and 3 has $\tau \in [1500, 1750]$. The
    last two columns are the mean value and the standard deviation
    computed using the values from the three different regions.
    \label{tab:xi}}
\end{table}

Analysing the comoving monopole separation, $\xim$, we can see that it
increases slower than $\tau$. However, it keeps increasing during the
whole evolution of the system, showing that $N$ decreases and that
pole-antipole annihilations are present in all the stages of the
evolution.

\subsection{Linear pole Density}

We can characterise the linear pole density (the number of poles
of a particular type per unit length of string) in two
different ways: $r$, the number per unit physical length in units of
the pole acceleration scale $1/\dBV$ (\ref{e:rDef}), and $n$, the number
per comoving string length.

\begin{figure} 
   \centering
   \includegraphics[width=3.5in]{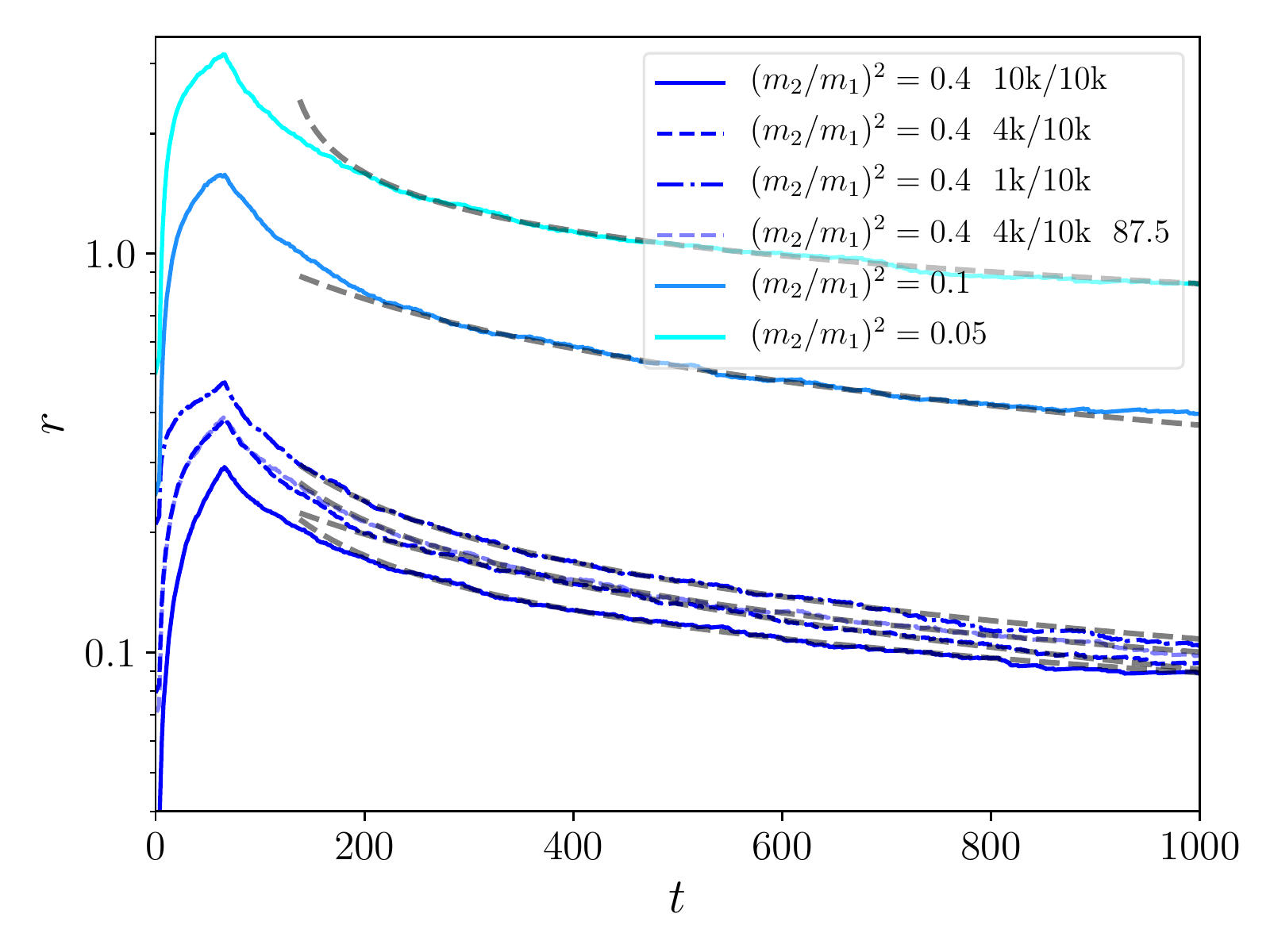} 
   \includegraphics[width=3.5in]{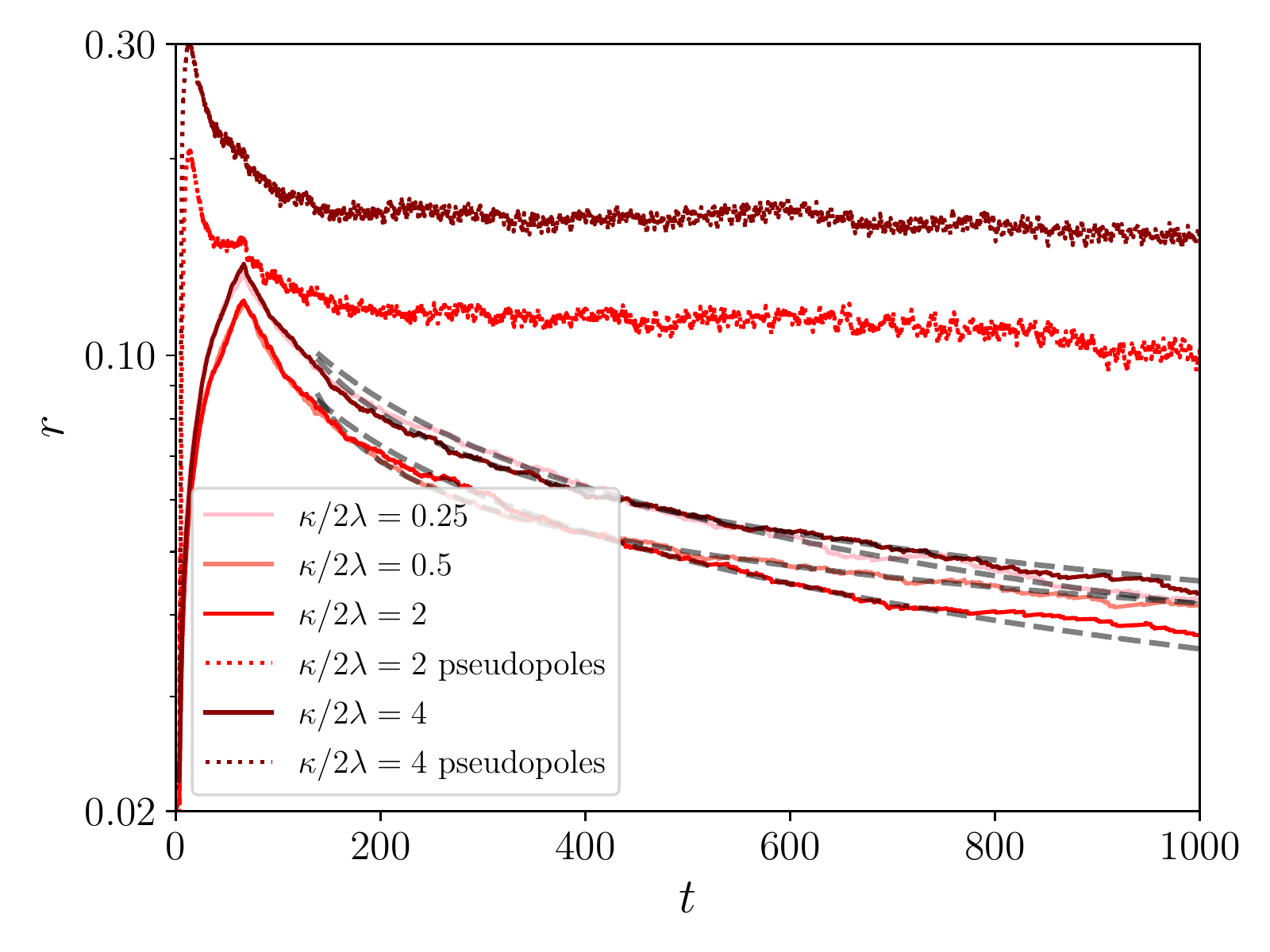} 
   \caption{Linear pole density $r$ 
     (\ref{e:rDef}) for necklaces with monopoles (top) and semipoles
     (bottom), plotted against physical time $t$.  The dashed grey
     line represents the fit to the data using the function presented
     in Eq.~(\ref{eq:rfit}).  The values for the fit parameters can be
     seen in Table~\ref{tab:r}. See the caption to
     Fig.~\ref{fig:xis} for an explanation of the legend. Note that in
     the plot for necklaces with semipoles we show also the linear
     pseudopole density in the cases where $\kappa/2\lambda>1$.}
   \label{fig:r}
\end{figure}

The ratio of pole to string energy density, $r$, is plotted in
Fig.~\ref{fig:r} against physical time, which for the radiation
dominated era is
\begin{equation}
t = \frac12 a(\tau) \tau.
\end{equation}
All the different cases simulated can be found in these figures. 
We can see that in all the cases the value of $r$ does not increase, once the physical evolution begins at $\tauCG$.

Also plotted is the number per unit physical length of pseudopoles. In
this case, $r$ seems to asymptote to a constant of order $10^{-1}$,
indicating a constant physical separation along the string. Although
there is no extra energy density associated with a pseudopole, it does
suggest that there is a physical length scale on the string of around
$10\dBV$ imprinted in the fields.

In order to analyse the power law with which $r$ decreases we fit with
the following function:
\begin{equation}
r = r_{b} [(t - t_0)/(t_{b}-t_0)]^{-\beta},
\label{eq:rfit}
\end{equation}
where $r_{b}$, $t_0$ and $\beta$ are the fitting parameters and we
choose $t_{b}$ to be the end of the fitting range $[t_{a}, t_{b}] =
[245, 750]$. The values of the fitting parameters can be found in
Table~\ref{tab:r}.

\begin{table}
\centering
\begin{tabular}{ccccr}
$m_2^2/m_1^2$ &$\kappa/2\lambda$ & $\beta$ & $r_{b}$ & $t_0$   \\ \hline
0.4 & 1   & $0.36 \pm 0.15$ & $0.10 \pm 0.01$ &  $60   \pm 3 $ \\
0.4 & 1   & $0.70 \pm 0.15$ &  $0.11 \pm  0.01$&  $-184 \pm 7$  \\
0.4 & 1   & $0.46 \pm 0.15$ &  $0.12 \pm 0.01$ & $28  \pm  3 $ \\
0.4 & 1   &  $0.42 \pm 0.15$ &  $0.11 \pm  0.01$ & $47 \pm 3$ \\ \hline
0.1 & 1   &  $0.57 \pm 0.15$ & $0.43 \pm 0.01$ & $-98 \pm 7$  \\
0.05 & 1 & $0.27 \pm 0.15$ & $0.92 \pm 0.01$ & $122 \pm 3$   \\ \hline
1 & 0.25 &  $0.46 \pm 0.15$ & $0.05 \pm 0.01$ & $-6 \pm 5$\\
1 & 0.5   &  $0.22 \pm 0.15$ & $0.04 \pm 0.01$ & $106 \pm 2$ \\
1 & 2      &  $0.46 \pm 0.15$ & $0.04 \pm 0.01$ & $-9 \pm  4$  \\
1 & 4      &  $0.30 \pm 0.15$ & $0.05 \pm  0.01$ & $ 68 \pm 3$ \\
\hline
\end{tabular}
\caption{Parameters computed from fitting $r$ in the range $t\in[245,
    750]$ using the function presented in Eq.~(\ref{eq:rfit}). The
  uncertainties for $\beta$ and $r_{b}$ are obtained using the
  variations in the values for the four necklace cases with the same
  physical parameters. However, the uncertainties for $t_0$ are
  obtained from the fitting because the value for $t_0$ can vary in
  simulations with the same physical parameters but different initial
  conditions.}
\label{tab:r}
\end{table}

\begin{figure} 
   \centering
   \includegraphics[width=3.5in]{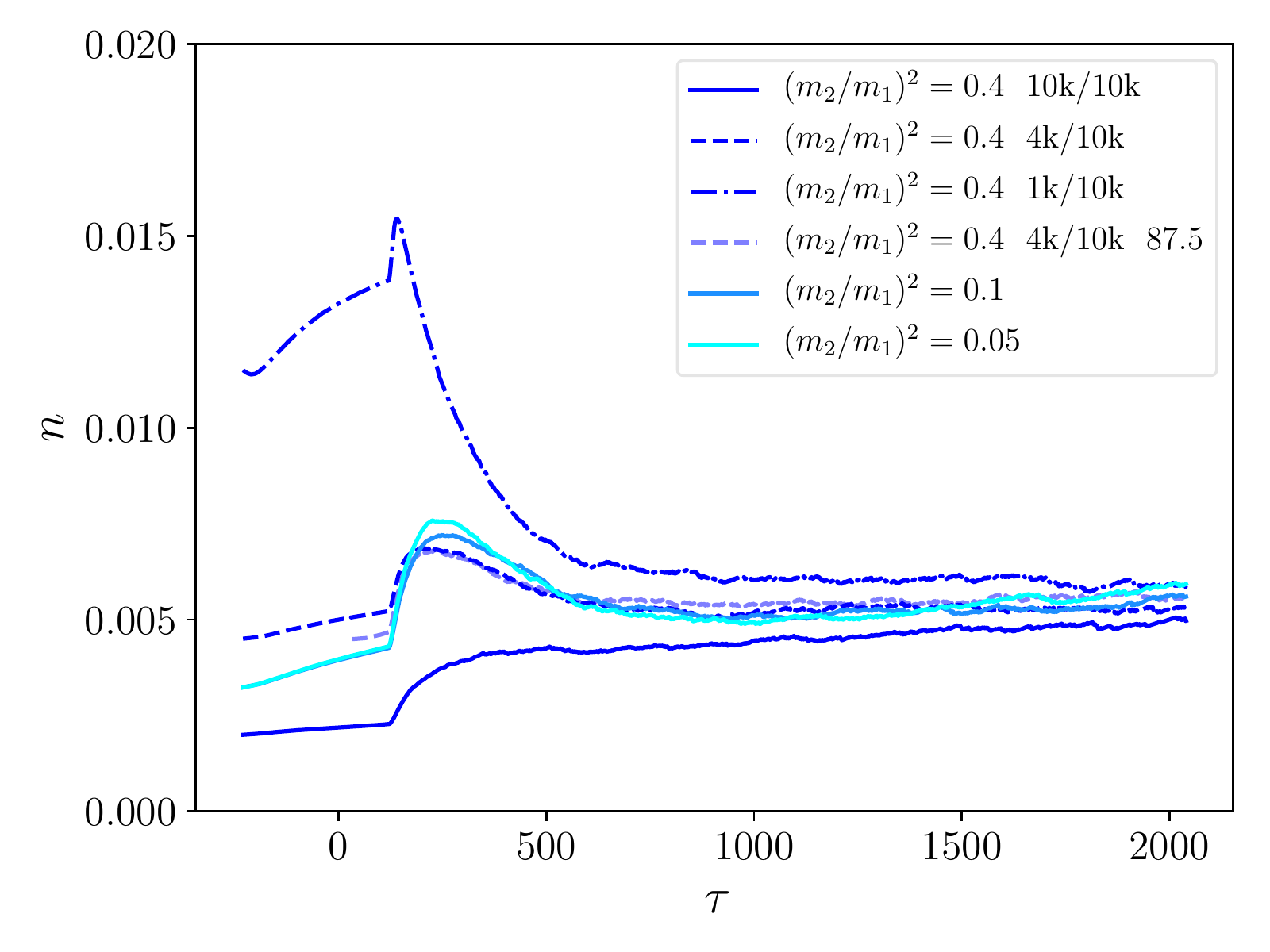} 
   \includegraphics[width=3.5in]{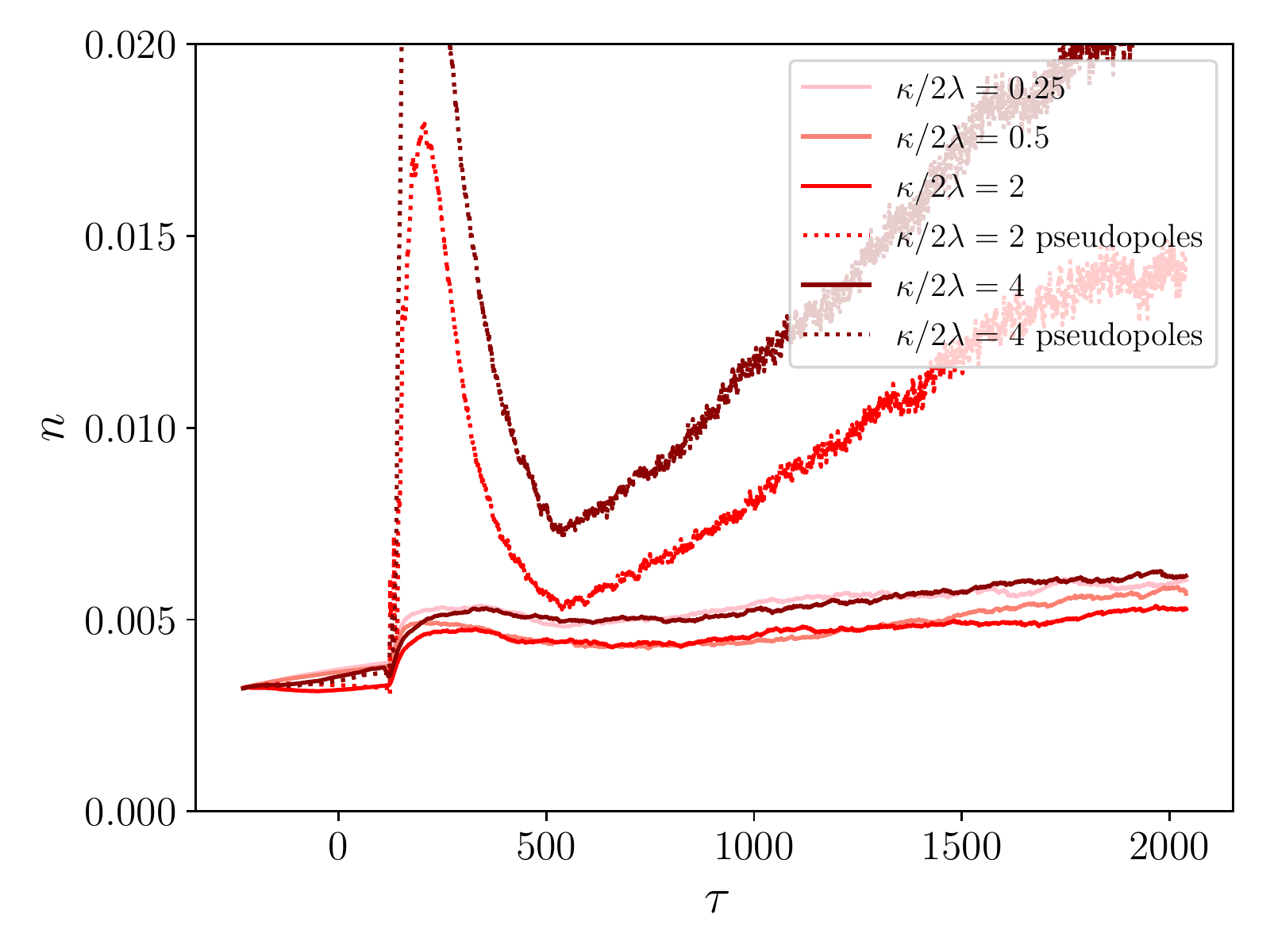} 
   \caption{The number of poles per comoving string length $n$ for the
     necklaces with monopoles (top) and semipoles (bottom), plotted
     against conformal time $\tau$; in the semipole case only one
       type of semipoles ($\mathbf{B}^{(1)}$ or $\mathbf{B}^{(+)}$) is
       shown. See the caption to Fig.~\ref{fig:xis} for an
     explanation of the legend. Note that in the plot for necklaces
     with semipoles, we show also the number of $\mathbf{B}^{(1)}$
     pseudopoles per comoving string length in the cases where
     $\kappa/2\lambda>1$.}
   \label{fig:n}
\end{figure}

The fits indicate that $r$ decreases with a power law close to
$t^{-1/2}$, which would indicate that the comoving density
$n=ar/d_{BV}$ should be approximately constant.  In Fig.~\ref{fig:n}
we can see that for $n$ does indeed appear to tend to a constant at
large time, consistent with the results in
Ref.~\cite{Hindmarsh:2016dha}

The asymptotic values of $n$ at large conformal time are about a
factor of 2 smaller than in our previous simulations, which has no
particular physical significance.  Instead, we note that the behaviour
$r \propto t^{-1/2}$ brings in a new length scale $D$, which can be
defined from
\begin{equation}
\label{e:Ddef}
r = \frac{\dBV}{\sqrt{2Dt}}.  
\end{equation}
Using the value of $n$ at the last time step of the simulation one can
obtain an approximate value for $D$. As we have already noted, all the
cases seem to asymptote to the same value of $n$, so we can extract an
estimate of a universal $D$ by taking the average over all the
realisations.  The value computed is $D=16 \pm 2$.

The results for $n$ for semipoles in Ref.~\cite{Hindmarsh:2016dha}
were not conclusive, and we now understand that in using the source of
${\bf B}^{(1)}$ flux to locate the semipoles was incorrect.  Our new
results for semipoles establish that they behave in the same way as
monopoles.

The linear density of the sources of ${\bf B}^{(1)}$ flux
(pseudopoles) is nonetheless instructive.  We have therefore also
plotted the pseudopole separation in Figs.~\ref{fig:r} and
\ref{fig:n}, for which $r$ asymptotes to O($10^{-1}$), and $n$
increases linearly with conformal time, as expected for an
asymptotically constant $r$.

\subsection{Velocities}

\begin{figure} 
   \centering
   \includegraphics[width=3.5in]{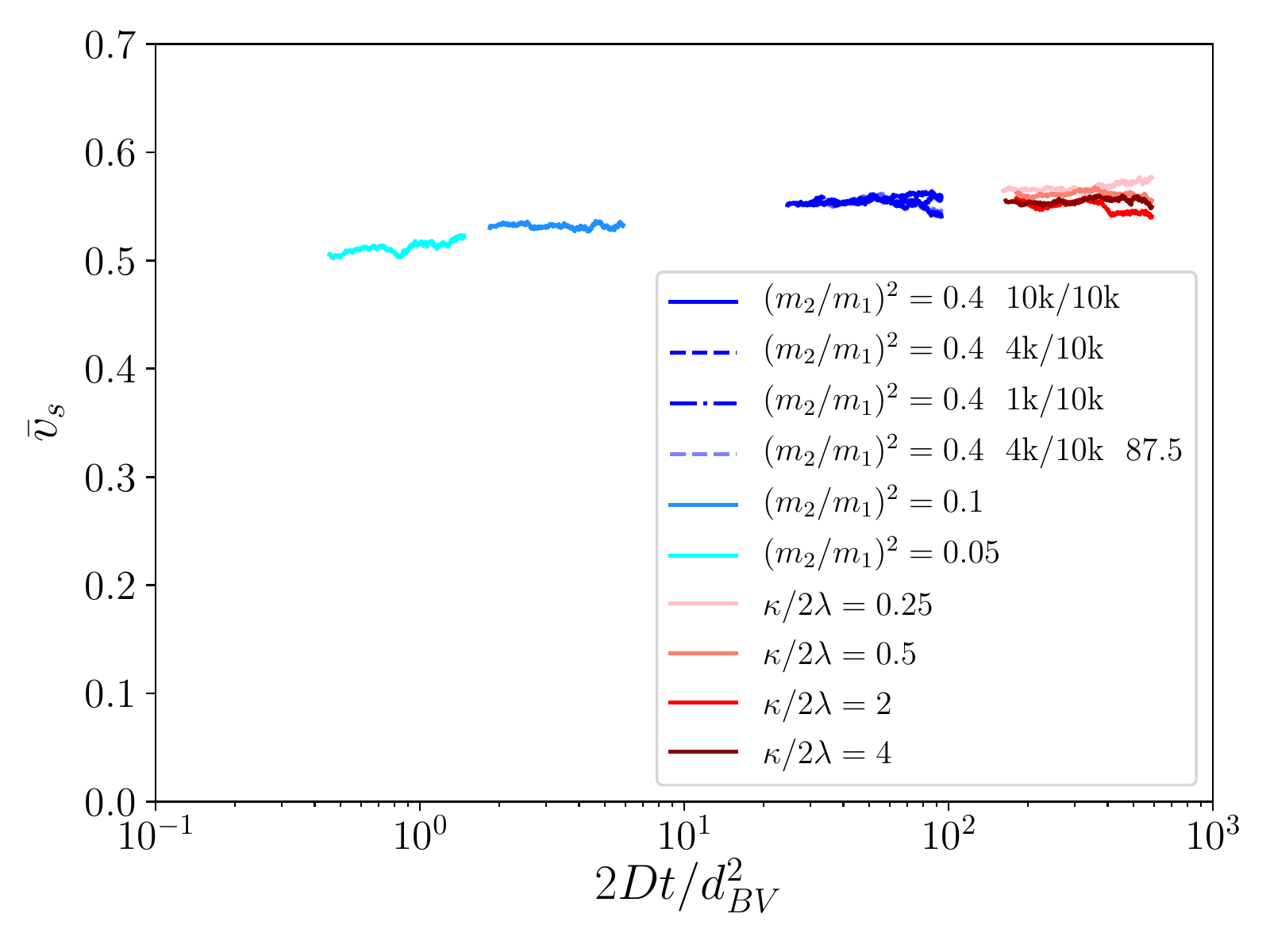} 
   \includegraphics[width=3.5in]{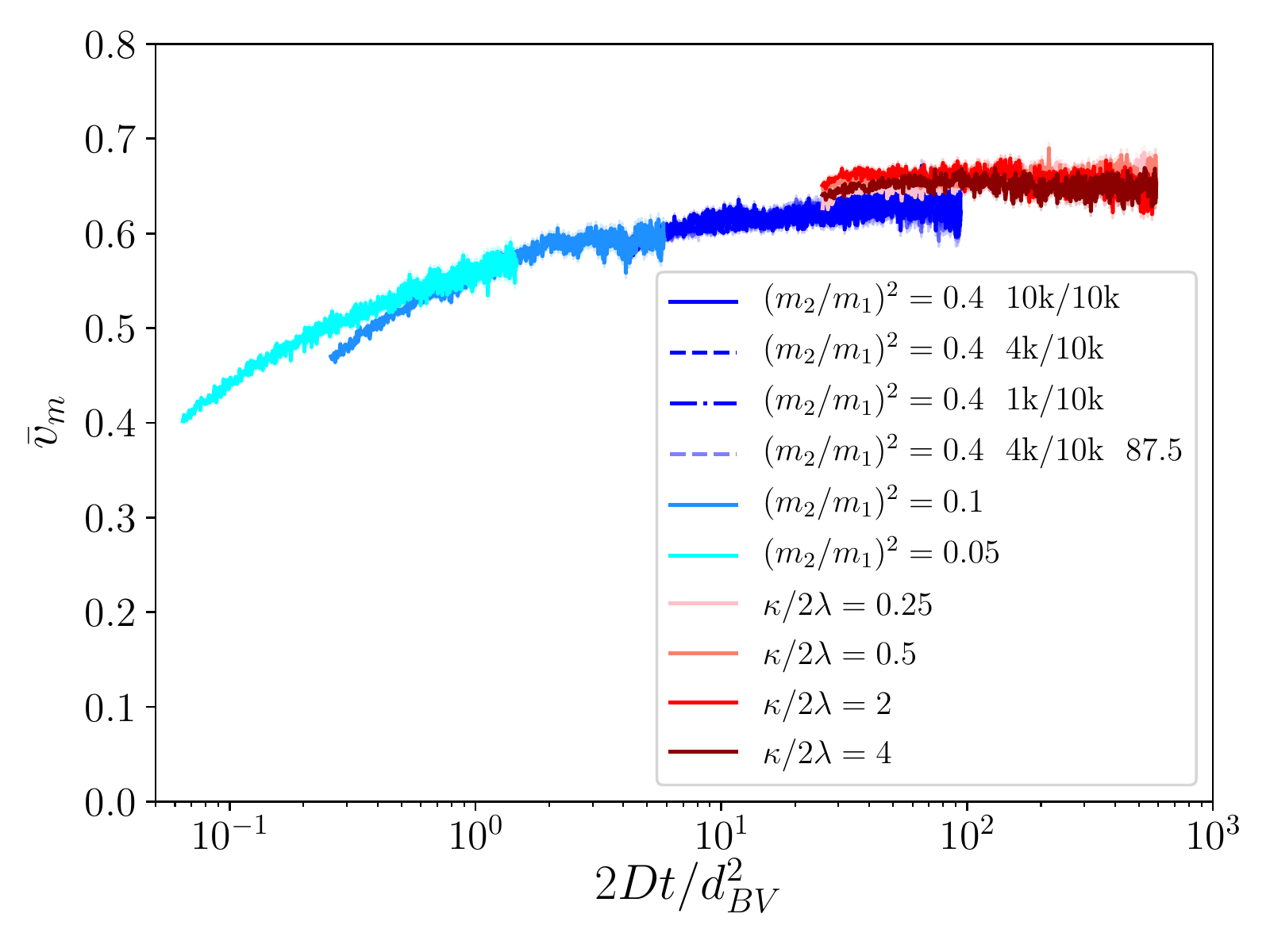} 
   \caption{The root mean square velocity for strings (top) and
     monopoles (bottom) computed by the method outlined in Section
     \ref{sec:measure}, plotted against $2Dt/\dBV^2$, where $t$ is
     physical time, $\dBV$ is the acceleration time scale
     (\ref{e:dBVdef}), and $D$ is the length scale defined from
     Eq.~(\ref{e:Ddef}).  The average values for the velocities can be
     found in Table~\ref{tab:vel}. See the caption to
     Fig.~\ref{fig:xis} for an explanation of the legend.  }
   \label{fig:v}
\end{figure}

In Fig.~\ref{fig:v}, we show the RMS velocities computed for strings
and poles using the procedure described in
Section~\ref{sec:measure}, plotted against physical time in units of
$\dBV^2/2D$.  
The velocities in different simulations fall on an
approximately consistent curve which appears to asymptote to a
constant at large times.  Semipoles move faster (see
Table~\ref{tab:vel}).  The curve is particularly noticeable for the
light strings, which need more time to accelerate the monopoles to their 
asymptotic speed. RMS velocity values can be seen in Table~\ref{tab:vel}.

One can obtain an estimate of the velocities of the monopoles and
semipoles along the string using the string and pole RMS
velocities $\vrel^2=\vrmsm^2-\vrmss^2$, also given in
Table~\ref{tab:vel}.  Note that $\vrel^2 \sim \vrmss^2$ in all cases,
with larger relative velocities for semipoles.
In our previous simulations we were unable to measure the RMS
velocities well enough to gain an unambiguous non-zero value for the
motion of the poles along the strings.

\begin{table}
\centering
\begin{tabular}{ccccc}
$m_2^2/m_1^2$ &$\kappa/2\lambda$ & $\bar{v}_s$ & $\bar{v}_m$ &$\bar{v}_{rel}$  \\ \hline
0.4 & 1   & 0.552 $\pm$ 0.005& 0.63 $\pm$ 0.01 &  0.30 $\pm$   0.05 \\
0.4 & 1   &0.558 $\pm$ 0.003& 0.629 $\pm$ 0.009 & 0.29 $\pm$ 0.04   \\
0.4 & 1   & 0.555 $\pm$ 0.002& 0.629 $\pm$ 0.008 &  0.30 $\pm$  0.04   \\
0.4 & 1   & 0.553 $\pm$ 0.004& 0.63 $\pm$  0.01& 0.29 $\pm$ 0.04 \\ \hline
0.1 & 1   & 0.532 $\pm$ 0.002& 0.592 $\pm$ 0.008 & 0.26 $\pm$ 0.04  \\
0.05 & 1 & 0.513 $\pm$ 0.005& 0.56 $\pm$ 0.01& 0.21 $\pm$ 0.07  \\ \hline
1 & 0.25 & 0.568 $\pm$ 0.004& 0.658 $\pm$ 0.009 &0.33 $\pm$  0.04 \\
1 & 0.5   & 0.561 $\pm$ 0.002& 0.660 $\pm$ 0.009& 0.35 $\pm$ 0.03 \\
1 & 2      & 0.549 $\pm$  0.002& 0.652 $\pm$ 0.008 & 0.35 $\pm$ 0.03   \\
1 & 4      & 0.555 $\pm$ 0.002& 0.648 $\pm$ 0.008 & 0.33 $\pm$ 0.03\\
\hline
\end{tabular}
\caption{Values of the velocities of the strings and poles and the
  pole velocity relative to the string. The velocities are
  computed in $t \in [297, 900]$. 
  The error shown is the standard deviation obtained from
  averaging over all the timesteps.}
\label{tab:vel}
\end{table}

\section{Comparison to necklace evolution models \label{sec:ana}}

Having presented the results of our simulations, we compare our
findings to the analytical models presented in the literature. The
models all make assumptions about the system, and derive various
predictions, which differ between models.  We can test the validity of
the assumptions and the correctness of the predictions in light of our
new results.

The first model describing the evolution of the necklace network was
introduced by Berezinsky and Vilenkin (BV) in
Ref.~\cite{Berezinsky:1997td}.  The authors assumed that there is no
motion of monopoles along the strings and that monopole-antimonopole
annihilation is negligible.  They also argued that the typical
velocity of the strings and monopoles was
\begin{equation}
\label{eq:vs-BV}
\vrmss \sim \frac{1}{\sqrt{1+r}},
\end{equation}
based on considering the necklace to have an effective mass per unit
length $\mu_\text{eff} = \mu + \mMon N/aL$, while maintaining tension
$\mu$.  They presented the following differential equation for $r$, in
the regime where $r \ll 1$:
\begin{equation}
\frac{\dot{r}}{r}=-\frac{\kappa_s}{t}+\frac{\kappa_g}{t}.
\label{eq:r-BV}
\end{equation}
The first term on the right hand side describes string stretching due
to the expansion of the Universe, and has $\kappa_s=\gamma
(1-2\vrmss^2)$, where $\gamma = t \dot a/a = \nu/(1+\nu)$.  The second one models
the competing effect of strings shrinking due to energy loss, with
$\kappa_g \simeq 1$.  In Ref.~\cite{Berezinsky:1997td} the primary
energy loss channel was thought to be gravitational radiation, but the
field radiation observed in the numerical simulations of topological
defects (see also Ref.~\cite{Hindmarsh:2017qff}) will also have the
same effect.

Using the string velocities obtained in our work and the estimated
value for $\kappa_g$ from Ref.~\cite{Berezinsky:1997td}, the solution
to Eq.~(\ref{eq:r-BV}) has $r$ growing with a power of time close to 1.
Their conclusion was therefore that if $r$ is initially small, it will
grow.

Our results show the contrary: $r$ decreases in all the cases that we
considered (see Fig.~\ref{fig:r}). Our simulations show that the
number of monopoles $N$ decreases during the evolution of the system,
demonstrating that monopole-antimonopole annihilations are important.
Animations of network evolution (See
Ref.~\cite{vimeo:necklaces,vimeo:semipoles025,vimeo:semipoles2})
indicate that annihilations take place both on long strings and loops.

\begin{figure} 
   \centering
   \includegraphics[width=3.5in]{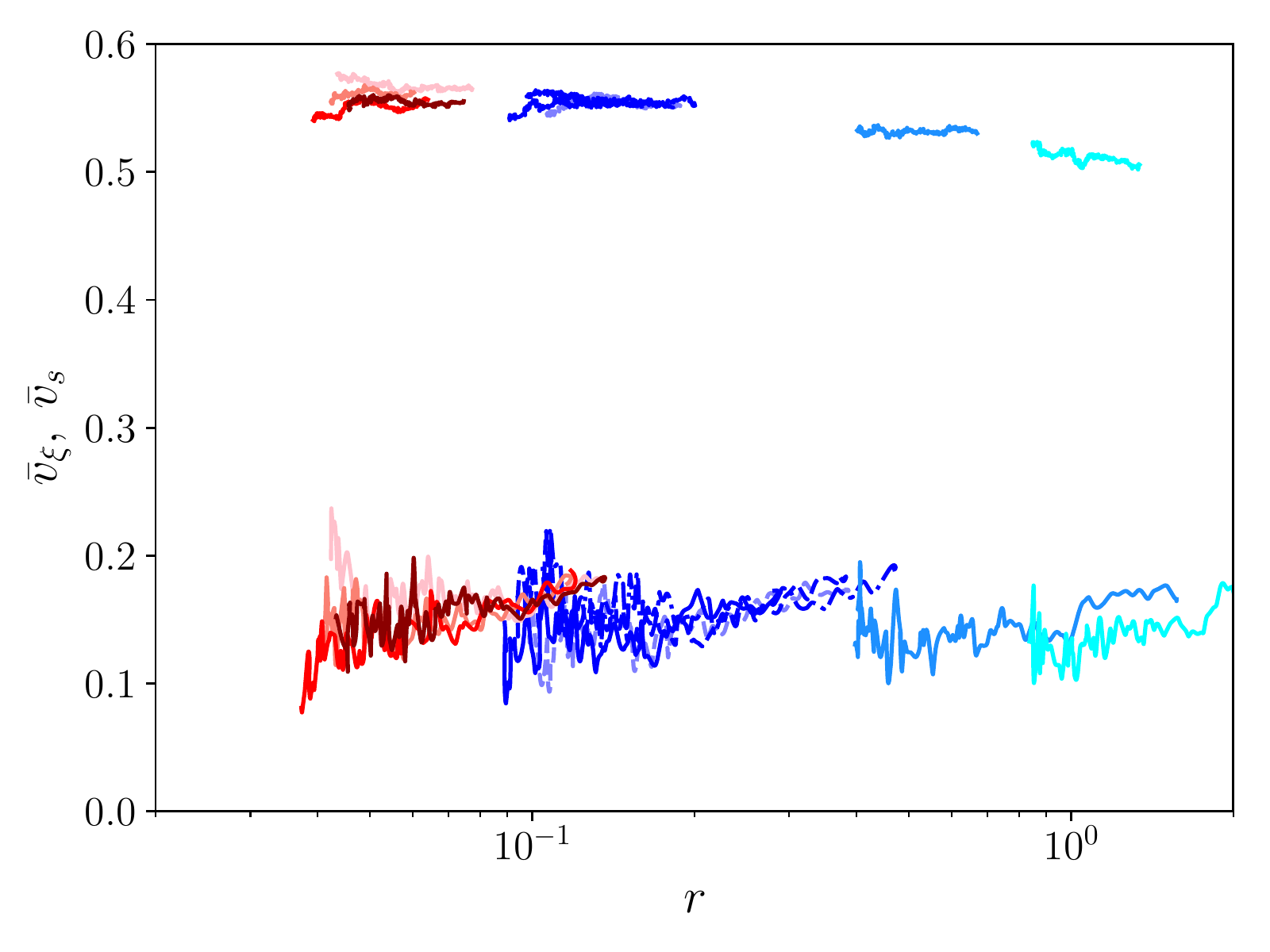} 
   \includegraphics[width=3.5in]{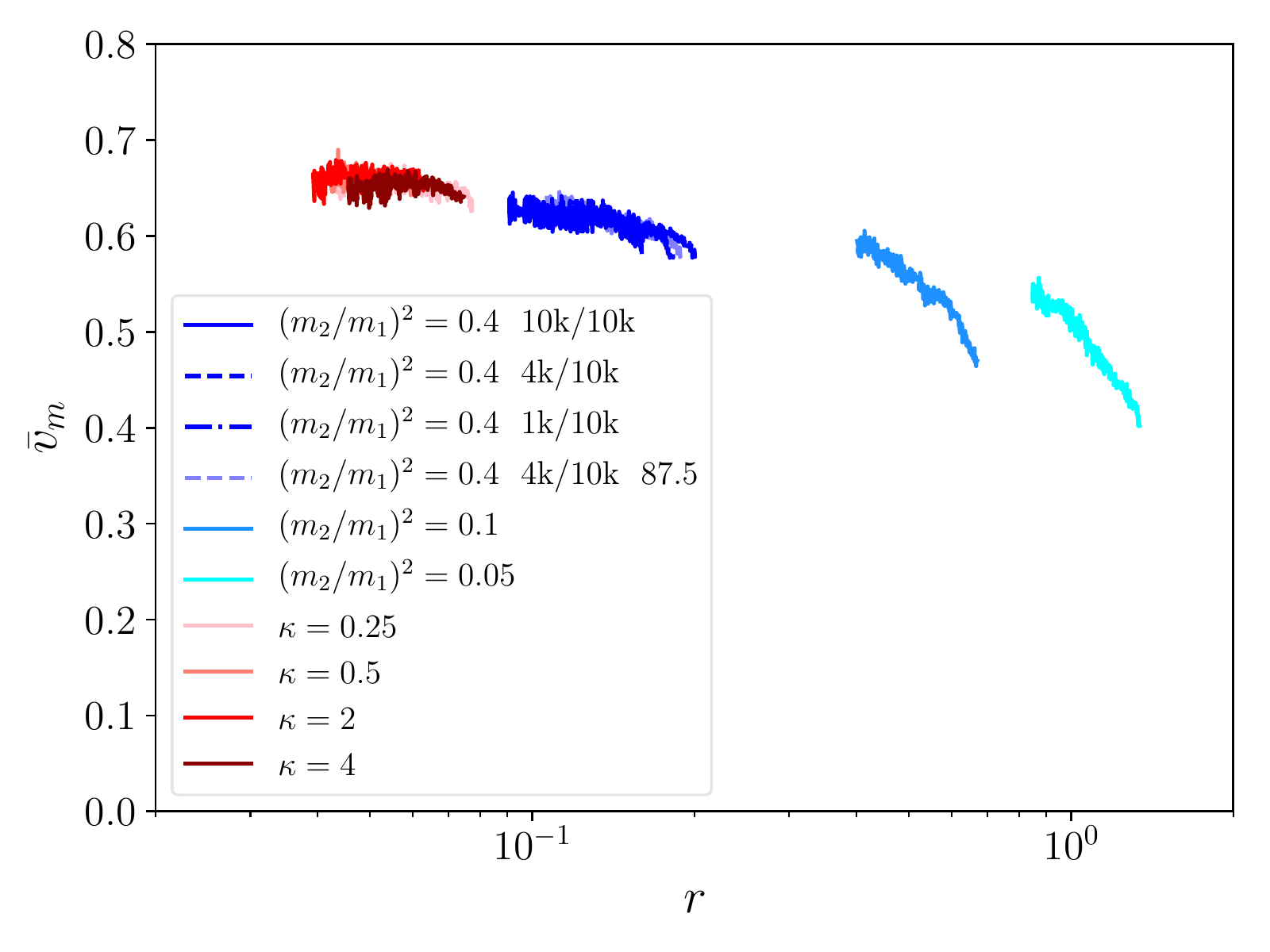} 
   \caption{Plots showing the RMS string velocity $\bar{v}_s$ against
     $r$ (top) and $\bar{v}_m$ against $r$ (bottom). In the top plot we have
     also included data showing $\bar{v}_\xi = d\xi_s/d\tau$ against
     $r$, where the $\bar{v}_\xi$ data is smoothed over a Blackman
     window with 101 points.  See the caption to Fig.~\ref{fig:xis}
     for an explanation of the legend, which is the same in both
     plots.  \label{fig:ragainstv} }
\end{figure}

Another major difference with Ref.~\cite{Berezinsky:1997td} is
in the dependence of the string velocity on $r$.  In
Fig.~\ref{fig:ragainstv} (top) we have plotted the directly computed
$\vrmss$ against $r$.  The gradient of the mean string
separation $d\xis/d\tau$ has dimensions of velocity, and provides an
estimate of the string RMS velocity on the scale $\xis$, which we
denote $\bar{v}_\xi$.  We have therefore also plotted $\bar{v}_\xi$,
smoothed with a Blackman filter over 101 time steps, which is clearly
distinguished by having much smaller values.

It is clear that there is no evidence for a dependence of the
large-scale velocity $\bar{v}_\xi$ on on $r$.  The short-distance
measure $\vrmss$ decreases very slightly for $r \simeq1$, but
certainly not by a factor $1/\sqrt{2}$ as predicted by
Eq.~(\ref{eq:vs-BV}).

The directly-computed monopole RMS velocity $\vrmsm$ is shown in
Fig.~\ref{fig:ragainstv} (bottom).  There is some evidence for a slow
decrease of the monopole RMS velocity with increasing $r$, which is
probably due to the correlation between higher $r$ and earlier times,
before the monopoles have picked up full speed.  By eye, there is some
suggestion that there is a common asymptote of $\vrmsm \simeq 0.7$ as
$r \to 0$, which is the long-time limit of the necklace evolution.

Monopole annihilation is incorporated into the model of Blanco-Pillado
and Olum~\cite{BlancoPillado:2007zr}. They also used the BV assumption
for the string velocities (\ref{eq:vs-BV}), but argued that there
should be approximate equipartition between the components of the
monopole and string RMS velocities, and therefore the RMS velocity
component of monopoles along the strings should be
\begin{equation}
\vrel^2 \simeq \vrmss^2/2.
\label{eq:relv}
\end{equation}
They concluded that there should be frequent encounters between
monopoles and anti-monopoles on the string, which would result in
efficient annihilations. The mean monopole spacing should therefore be
of order $t/\vrel$ (physical units), and hence $r$ should decrease as
$t^{-1}$.

Our results are consistent with approximate velocity equipartition,
$\vrel \sim \vrmss$ (see Table~\ref{tab:vel}). However, our results
for $r$ are inconsistent with the $r  \propto t^{-1}$ behaviour 
predicted in Ref.~\cite{BlancoPillado:2007zr}.
For monopoles and semipoles, the fits to a power
law are closer to $r \propto t^{-1/2}$, consistent with constant
comoving linear density $n$. (See Table~\ref{tab:r})

The third model is a velocity-dependent one-scale model for monopoles
\cite{Martins:2008zz} adapted for the evolution of necklaces
\cite{Martins:2010ma}. This model focuses on the evolution of the
separation between monopoles, assuming that the string velocity obeys
Eq.~(\ref{eq:vs-BV}) and that the mean string separation is
similar to the mean monopole separation,
\begin{equation}
\label{e:MAssA}
\xim \sim \xis .
\end{equation}
With these assumptions for the strings, it should be sufficient to
study the mean separation and RMS velocity of the monopoles, and the
evolution equations for these parameters were derived to be (in our
notation)
\begin{eqnarray}
3\frac{d\xim}{dt} &=&(3+\vrmsm^2)H \xim+Q_*, \label{eq:vos1} \\
\frac{d\vrmsm}{dt} &=& (1-\vrmsm^2)\left(\frac{k_s}{ \dBV}-H\vrmsm\right),
\label{eq:vos2}
\end{eqnarray}
where $H$ is the Hubble parameter, $k_s$ the phenomenological string
curvature parameter \cite{Martins:1996jp,Martins:2000cs} and $Q_*$ a
constant energy loss term.

The solution of Eqs.~(\ref{eq:vos1},\ref{eq:vos2}) has $\xim \propto t$ and $\vrmsm
\rightarrow 1$.  This describes an evolution where $r \propto t^{-1}$
and the monopoles' Lorentz factor continually increases with time.
Again, this disagrees with our results, which indicate that $r \propto
t^{-1/2}$ and $\vrmsm \simeq 0.6$.

In conclusion, we can say that none of the models of which we are
aware describes our results: the key difference is the behaviour of
$r$, the linear physical monopole density in units of $\dBV$.  The
physical linear density decreases -- in contradiction to the BV model
-- as a result of monopole annihilation.  However, the monopole
annihilation cannot be as efficient as assumed in the other two
models, as $r$ decreases in proportion to $t^{-1/2}$ rather than
$t^{-1}$. Semipoles behave like monopoles.

\section{Discussion \label{sec:dis}}

We have carried out the largest simulations to date of systems of
necklaces, studying both monopoles and semipoles, exploring a wider
range of string-to-monopole energy density ratios $r$ than before, and
following the evolution to larger string separations.

Our results concern the mean comoving string separation $\xis$, the mean comoving monopole (or semipole) separation $\xim$, the mean RMS string velocity $\vrmss$, and the mean RMS monopole velocity $\vrmsm$.

The mean comoving string separation $\xis$ always increases with
conformal time, consistent with linear scaling $\xis \propto \tau$.
The slopes are shown in Table~\ref{tab:xi}.  The mean separation of
monopoles and semipoles, grows as $\xim \propto \tau^{2/3}$. The rest
have $r$ decreasing in proportion to $t^{-1/2}$, equivalent to a a
constant comoving linear density $n$ (see Fig.~\ref{fig:n}).  In terms
of the physical mean separation and physical time,
$\xi_\text{m}^\text{phy} \propto t^{5/6}$.

String RMS velocities tend to a constant value $\vrmss \simeq 0.55$,
only weakly dependent on the string-to-monopole energy density ratio
$r$.

Monopole and semipole RMS velocities evolve slowly towards a constant
value around $0.7$ at the end of our simulations, on a
timescale controlled by the monopole acceleration parameter $1/\dBV$.
The RMS velocities in the limit of vanishing string-to-monopole energy
density ratio $r$ appear to be tending to a common value around $0.7$.

Models of necklace evolution in the literature do not describe our
results.  A key point is that the assumed dependence of the RMS string
velocity on the monopole-to-string density ratio $r$ (\ref{eq:vs-BV})
is not observed.  Instead, the RMS string velocity barely depends on
$r$ at all, up to $r \simeq 2$.  Thus the picture of massive monopoles
as slowing down the strings is incorrect; instead, it seems that the
strings can drag the monopoles around with them, although the more
massive the monopoles, the longer it takes for their RMS velocity to
reach that of the strings.

Monopole and semipole annihilation is certainly important, contrary to
\cite{Berezinsky:1997td}, but has much lower efficiency than envisaged
in Ref.~\cite{BlancoPillado:2007zr}, who argued that monopoles would
annihilate with probability of order unity if they encountered each
other on the string.  If the poles have an RMS velocity along the
string of $\vperp$, the average pole should encounter others at a
conformal time rate $\vperp n$. Thus if $\si$ is the annihilation
probability, we should be able to write a one-dimensional Boltzmann
equation
\begin{equation}
  \label{e:BolEqn}
\frac{dn}{d\tau} = n \left(- \si \vperp n + 2 \frac{1}{\xis} \frac{d
  \xis}{d\tau} \right),
\end{equation}
where the second term on the right hand side describes the increase in
the comoving linear density due to the string shrinking.  It seems
reasonable at first sight to identify $\vperp$ with $\vrel$, and
assuming constant $\si$, this equation would have a solution $ n =
\nu_0/\tau$, with $\nu_0 = 3/\si\vrel$. This is equivalent to $r
\propto t^{-1}$, and is essentially the model put forward in
Ref.~\cite{BlancoPillado:2007zr}.  The fact that $n$ appears to tend
to a constant is inconsistent with the model, and therefore at least
one of the assumptions that go into it.  Either there is some
mechanism suppressing annihilation, or it is incorrect to make the
identification $\vperp \sim \vrel$.

The constraint that semipoles can annihilate only with a corresponding
anti-semipole does not appear to significantly change their
annihilation rate in comparison to monopoles.

We do not have a clear idea of how the suppression of pole
annihilation happens, despite their appreciable short-distance motion
along the string, $\vrel \simeq 0.3$.  One possibility is that $\vrel$
is a short-distance measure of velocity, while $\vperp$ is effectively
averaged over a scale $d$, the average comoving separation of poles
along the string.  This measure of velocity could decrease as
$\tau^{-1}$ if the pole motion were more like diffusion than uniform
linear translation.  Perhaps short distance fluctuations on the
string, analogous to the L\"uscher term on the QCD string
\cite{Luscher:1980ac}, act to keep the monopoles in some kind of
Brownian motion.

As explained earlier, $r \propto t^{-1/2}$ brings in a new length
scale $D$, which can be defined from $r = \dBV/\sqrt{2Dt}$, indicating
that the RMS linear separation between poles is $\sqrt{2Dt}$.  This
could be explained by the poles executing Brownian motion, with
diffusion constant $ D \simeq 16$ in lattice units, and annihilating
with O(1) probability when meeting. The average velocity on the pole
separation scale would go as $\sqrt{2D/t}$, proportional to
$\tau^{-1}$ as required for the constant $n$ solution to
(\ref{e:BolEqn}). We do not have a good microscopic explanation for
the value of $D$, although we note an order-of-magnitude coincidence
with the separation of pseudopoles, sources of a certain U(1) flux not
associated with a local increase of energy density.  Significant
computer time would be required to investigate pole annihilation
further.

In summary, we have found strong evidence that the necklace network as
a whole scales, in the sense that its energy density remains a
constant fraction of the total energy density, now for semipoles as
well as monopoles \cite{Hindmarsh:2016dha}.  The fractional energy
density of poles decreases as $t^{-1/2}$, suggesting a diffusive
process.  The energy in the necklaces is lost to radiative modes of
the gauge and scalar fields.

The cosmological implications of this kind of scaling necklace network
were discussed in Ref.~\cite{Hindmarsh:2016dha}; in summary, the
principal observational constraints come from diffuse $\gamma$-rays
for necklaces in a sector with substantial couplings to the Standard
Model ($G\mu \lesssim 3 \times 10^{-11}$) or the Cosmic Microwave
Background for necklaces in a hidden sector ($G\mu \lesssim 10^{-7}
$).

\begin{acknowledgments}
MH (ORCID ID 0000-0002-9307-437X) acknowledges support from the
Science and Technology Facilities Council (grant number
ST/L000504/1). ALE (ORCID ID 0000-0002-1696-3579) is grateful to the
Early Universe Cosmology group (Basque Government grant IT-979-16) of
the University of the Basque Country for their generous hospitality
and useful discussions. DJW (ORCID ID 0000-0001-6986-0517) and AK
(ORCID ID 0000-0002-0309-3471) acknowledge support from the Research
Funds of the University of Helsinki. The work of DJW was performed in
part at the Aspen Center for Physics, which is supported by National
Science Foundation grant PHY-1607611. This work is supported by the
Academy of Finland grant 286769.  We are grateful to Kari Rummukainen
for many useful discussions, and particular for alerting us to the
L\"uscher term. The simulations for this paper were carried out at the
Finnish Centre for Scientific Computing CSC.
\end{acknowledgments}

\bibliography{scone} 

\end{document}